\documentstyle[amsfonts,epsfig,11pt]{article}

\def\mypagenumber{1}

\def\myend{\end{document}}

\normalsize

\newcounter{sxn}

\newcounter{axn}

\date{}

\newdimen\mybaselineskip
\newcommand{\beeq}{\begin{equation}}
\newcommand{\eneq}{\end{equation}}
\newcommand{\be}{\begin{eqnarray}}
\newcommand{\ee}{\end{eqnarray}}
\newcommand{\bpic}{\begin{picture}}
\newcommand{\epic}{\end{picture}}

\def\la{\raise.16ex\hbox{$\langle$} \, }
\def\ra{\, \raise.16ex\hbox{$\rangle$} }

\def\psibar{ \psi \kern-.65em\raise.6em\hbox{$-$} }
\def\mbar{ m \kern-.78em\raise.4em\hbox{$-$}\lower.4em\hbox{} }

\def\n@space{\nulldelimiterspace=0pt \mathsurround=0pt }
\def\huge#1{{\hbox{$\left#1\vbox to 20.5pt{}\right.\n@space$}}}

\def\myskip{\noalign{\kern 8pt}}
\def\myeqspace{\noalign{\kern 10pt}}

\def\boxit#1{$\vcenter{\hrule\hbox{\vrule\kern3pt
    \vbox{\kern3pt\hbox{#1}\kern3pt}\kern3pt\vrule}\hrule}$}
\def\bigbox#1{$\vcenter{\hrule\hbox{\vrule\kern5pt
     \vbox{\kern5pt\hbox{#1}\kern5pt}\kern5pt\vrule}\hrule}$}

\def\ignore#1{{}}

\begin{document}

\bibliographystyle{unsrt}
\footskip 1.0cm

\thispagestyle{empty}
\setcounter{page}{\mypagenumber}


\begin{flushright}{BRX-TH-521\\}
\end{flushright}

\vspace{2.5cm}
\begin{center}
{\LARGE \bf {Energy in Topologically Massive Gravity }}\\
\vspace{2cm}
{\large S. Deser \hskip 0.3 cm and \hskip 0.3 cm
Bayram Tekin\footnote{e-mail:~
deser,tekin@brandeis.edu}~\footnote{Address after July 15,
{\it{Physics Department, Middle
East Technical University, 06531 Ankara, Turkey}.} }}\\
\vspace{.5cm}
{\it {Department of Physics, Brandeis University, Waltham, MA 02454,
USA}}\\

\end{center}

\vspace*{1.5cm}


\begin{abstract}
\baselineskip=18pt
We define conserved gravitational charges
in -cosmologically extended- topologically massive gravity ,
exhibit them in surface integral form about their de-Sitter or 
flat vacua and verify their correctness in terms of two basic 
types of solution.
\end{abstract}
\vfill


\newpage



\normalsize
\baselineskip=24pt plus 1pt minus 1pt
\parindent=25pt

Recently \cite{deser1}, we defined and computed the conserved
charges (particularly energy) for generic higher curvature gravity
models in surface integral, flux, form about the relevant,
asymptotically constant or zero curvature vacua. Our work involved the
formalism of \cite{ad} about asymptotically (Anti)-de-Sitter (AdS)
backgrounds in cosmological Einstein gravity. Here, we define
and compute the energy in the rather different context 
of topologically 
massive gravity (TMG) \cite{deser2} extended to include a 
cosmological term \cite{des} 
[Energy was defined in the original, $\Lambda
=0$, TMG  \cite{deser2}, but not in its surface integral form in 
terms of generic background Killing vectors.]

We begin directly with the vacuum equations \be R_{\mu\nu} \,-
{1\over 2} g_{\mu\nu}R +\Lambda g_{\mu \nu} + {1\over \mu }C_{\mu
\nu}= 0, \label{eom1} \ee \be C^{\mu\nu} \equiv {1\over
\sqrt{-g}}\epsilon^{\mu\alpha\beta} \nabla_\alpha (
R^{\nu}\,_\beta - {1\over 4}\delta^\nu\,_\beta R). \ee The Cotton
tensor $C^{\mu \nu}$ is symmetric, traceless and identically
conserved, while the parameter $\mu$ is the mass of the linearized
TMG excitations at $\Lambda=0$. $C^{\mu \nu}$, being the $D=3$
conformal curvature tensor, vanishes for any Einstein space,
including all external solutions of the cosmological Einstein
equations, such as AdS, Schwarzschild-dS and BTZ black holes.
More characteristic are geometries 
that obey the full TMG equations but not
their Einstein part alone, some of which are given in 
\cite{vurio,nutku, clement}, but these spaces do not correspond to 
bounded distributions.
There is as yet no known``Schwarzschild '' 
solution, let alone more complicated asymptotically de-Sitter 
or flat (for $\Lambda =0$ ) ones. What has been solved 
explicitly is the linearized metric generated by massive 
spinning interiors in the $\Lambda =0 $ sector \cite{deseranyon}, 
which will prove useful in testing our asymptotic expressions 
for the conserved generators.

We consider asymptotically AdS metrics ~\footnote{Our conventions
are: signature $(-,+,+)$,
$[\nabla_\mu, \nabla_\nu]V_{\lambda} =
R_{\mu \nu \lambda}\,^\sigma V_\sigma $,\,\,
$ R_{\mu \nu} \equiv R_{\mu \lambda \nu}\,^\lambda$.}
\be
g_{\mu \nu} \equiv  \bar{g}_{\mu \nu} + h_{\mu \nu},
\ee
where $h_{\mu \nu}$ is a (finite) deviation about the
background $\bar{g}_{\mu \nu}$ that obeys 
\be
\bar{R}_{\mu \lambda \nu \beta } =  \Lambda ( \bar{g}_{\mu
\nu} \bar{g}_{\lambda \beta} - \bar{g}_{\mu \beta}
\bar{g}_{\lambda \nu } ), \hskip 0.5 cm \bar{R}_{\mu \nu}= 2\Lambda
\bar{g}_{\mu \nu}, \hskip 0.5 cm \bar{R} = 6\Lambda.
\label{background}
\ee
Next, we expand the field equations
(\ref{eom1}) about $\bar{g}_{\mu \nu}$; as usual, the
nonlinear part is the energy momentum tensor (including matter if present):
\be
{\cal{G}}^L\,_{\mu \nu} + {1\over \mu } C^L\,_{\mu \nu} \equiv
\kappa T_{\mu \nu },
\ee
where the linear cosmological Einstein and Cotton terms read respectively,
\be
&&{\cal{G}}^L\,_{\mu \nu} \equiv  R^L\,_{\mu\nu} - {1\over 2}
\bar{g}_{\mu \nu}R_L - 2\Lambda  h_{\mu\nu},
\label{lineareinstein} \\
&& C_L\,^{\mu \nu}= {1\over \sqrt{-\bar{g}}}\epsilon^{\mu\alpha
\beta}\,\, \bar{g}_{\beta \sigma} \nabla_{\alpha}\,\left \{
{{R}}_L\,^{\sigma \nu} - 2\Lambda h^{\sigma \nu} - {1\over 4}
\bar{g}^{\sigma \nu} R_L  \right \}. \label{linearcotton} \ee The
linear part of the Ricci tensor is \be R^L_{\mu \nu} = {1\over 2}
\left \{ - \Box  h_{\mu\nu} - \nabla_\mu \nabla_\nu h  +
\nabla^{\sigma} \nabla_\nu h_{\sigma \mu} + \nabla^{\sigma}
\nabla_\mu h_{\sigma \nu} \right  \} \label{linearricci}. \ee Both
({\ref{lineareinstein}}) and (\ref{linearcotton}) are conserved,
 symmetric, and $\bar{g}^{\mu \nu}C^L\,_{\mu \nu}=0$ . To every
background Killing vector
 $\bar{\xi}^\mu$ corresponds a (background) conserved charge
\be Q^\mu(\bar{\xi}) &=& \int_{\cal{M}} d^2 x \sqrt{-\bar{g}}
T^{\mu \nu}\bar{\xi}_\nu = \int_{\cal{M}} d^2 x \sqrt{-\bar{g}}
\left \{ {\cal{G}}_L\,^{\mu \nu}\bar{\xi}_\nu + {1\over
\mu}C_L\,^{\mu \nu} \bar{\xi}_\nu \right \} \\ \nonumber &&\equiv
Q^\mu\,_{\mbox{E}} + Q^\mu\,_{\mbox{C}}  \label{charge}. \ee

Next we express (9) as a 1-dimensional surface integral
on the boundary. In the second paper of \cite{deser1}, we
gave a detailed account of how this is done for the Einstein part.
Here we simply quote that result and move on to the Cotton part.
\be
Q^{\mu}(\bar{\xi})\,_{\mbox{E}} = 
{1\over 8 \pi G}\int_{\partial{\cal{M}}} &dS_i&\Big \{
\bar{\xi}_\nu \bar{\nabla}^{\mu}h^{i \nu} -\bar{\xi}_\nu
\bar{\nabla}^{i}h^{\mu\nu}
+\bar{\xi}^\mu \bar{\nabla}^i h
-\bar{\xi}^i \bar{\nabla}^\mu h \nonumber \\
&&+h^{\mu \nu}\bar{\nabla}^i \bar{\xi}_\nu
- h^{i \nu}\bar{\nabla}^\mu \bar{\xi}_\nu +
\bar{\xi}^i \bar{\nabla}_{\nu}h^{\mu \nu}
-\bar{\xi}^\mu \bar{\nabla}_{\nu}h^{i \nu} + h\bar{\nabla}^\mu
\bar{\xi}^i \Big \},
\label{ad} 
\ee 
where $i$ denotes the space direction orthogonal to the boundary 
$\partial {\cal{M}}$.

To write $Q^\mu\,_{\mbox{C}}$ in (9) as a 
surface~\footnote{ We do not go into detail about the
complications of cosmological horizons for de-Sitter, as
opposed to AdS. For the former, the `boundary' we work with is
inside the cosmological horizon \cite{deser1}. }
integral, it is convenient to express $C_L\,^{\mu \nu}$ in
explicitly symmetric form : \be C_L\,^{\mu \nu} = { 1\over 2
\sqrt{-\bar{g}}} \left \{ \epsilon^{\mu
\alpha}\,_\beta\bar{\nabla}_\alpha {\cal{G}}_L\,^{\nu \beta} +
\epsilon^{\nu \alpha}\, _\beta \bar{\nabla}_\alpha
{\cal{G}}_L\,^{\mu \beta} \right \}. \ee Moving the Killing vector
inside the covariant derivatives, one obtains

\be 2 \bar{\xi}_\nu C_L\,^{\mu \nu}\sqrt{-\bar{g}} =
\bar{\nabla}_{\alpha}\left \{ \epsilon^{\mu \alpha \beta}
{\cal{G}}^L\,_{\nu \beta}\bar{\xi}^\nu +\epsilon^{\nu
\alpha}\,_\beta {\cal{G}}_L\,^{\mu \beta}\bar{\xi}_\nu
+\epsilon^{\mu \nu \beta} {\cal{G}}^{L\, \alpha} \,_{\beta}
\bar{\xi}_\nu \right \} + \epsilon^{\alpha \nu}\,_\beta
{\cal{G}}_L\,^{\mu \beta} \bar{\nabla}_\alpha \bar{\xi}_\nu
\label{cotsurface}. \ee The terms in the curly bracket are already
in the desired surface form. To put the last term into that form,
we can write it as $ X_\beta {\cal{G}}_L\,^{\mu \beta}$, where
$X^\beta = \epsilon^{\alpha \nu \beta}\bar{\nabla}_\alpha
\bar{\xi}_\nu $ is also a (background) Killing vector because $\xi$ is. 
This means that the last term is exactly like the Einstein part (\ref{ad}),
but in terms of $X^\mu $ instead of $\bar{\xi}^\mu$; hence we can immediately 
write it in surface form as well. Now we can compactly express 
the conserved charges in TMG, \be Q^\mu
 (\bar{\xi}) &=& Q^\mu\,_{\mbox{E}}(\bar{\xi})  + {1\over \mu}\oint
d S_i  \left \{ \epsilon^{\mu i \beta} {\cal{G}}^L\,_{\nu
\beta}\bar{\xi}^\nu +\epsilon^{\nu i}\,_\beta
{\cal{G}}_L\,^{\mu \beta}\bar{\xi}_\nu +\epsilon^{\mu \nu \beta}
{\cal{G}}^{L\, i}\,_\beta \bar{\xi}_\nu \right \}\\ \nonumber
&&+{1\over \mu} Q^\mu\,_{\mbox{E}}(\epsilon
\bar{\nabla}\bar{\xi}). \label{chargeF} \ee This is the desired
background-gauge invariant surface integral expression. 
The last term could be reduced further, but (13 ) is sufficiently simple. 

We may now compute the energy, {\it{i.e.}}, the charge corresponding to
the time-like AdS Killing vector for the two general classes of 
TMG solutions. First consider Einstein spaces, for which the Einstein and 
Cotton parts of TMG vanish separately. Here the prototype is the SdS 
metric; in static coordinates it is 
is \be ds^2 = -( 1- r_0 - \Lambda r^2 )dt^2
+ ( 1- r_0 - \Lambda r^2 )^{-1}dr^2 + r^2 d\varphi^2
\label{3dmetric}. \ee In this frame,the background, namely the
$r_0= 0$ version of (\ref{3dmetric}), has time-like Killing vector
$\bar{\xi}^\mu = (-1, {\bf{0}} )$. Then $X^\mu = ({\bf{0}},
2\Lambda r )$. Explicit computation shows that $Q^0_E( \epsilon
\bar{\nabla}\bar{\xi})$ in (13) vanishes. It is also
easy to see that the middle term does not contribute to the energy
for {\it{any}} Einstein space since  ${\cal{G}}_L\,^{\mu \nu} = 0
$. Thus, overall, SdS receives no contribution to its energy from
the Cotton tensor term. This was also observed in \cite{deser2} for the
corresponding asymptotically flat Schwarzschild solution. Finally,
the first term in (13) yields the energy $E$ to be proportional $r_0$ .

Next, we evaluate our charges for ``non-degenerate' solutions of TMG,
those not already obeying the Einstein equations. Here, the linearized
results of \cite{deseranyon} suffice because in the surface expressions 
at infinity, only these leading parts contribute. In a suitable gauge,
the metric exterior to bounded massive, spinning sources (for $\Lambda =0$ ) 
is given by
\be
h_{0i} = -\epsilon_{i j}\partial_j W(r), \hskip 1 cm 
h_{ij} = \phi(r)\delta_{ij}, \hskip 1 cm  h_{00} = n(r), 
\label{lineardes}
\ee 
where
\be
&&W(r) = -{1\over \mu }(m + \mu \sigma )( \log(r) +K_0(\mu r ) ), 
\hskip 1 cm \phi(r)= ( m + \mu \sigma )K_0(\mu r) + 2m\log(r), 
\nonumber \\
&& n(r) = ( m + \mu \sigma )K_0(\mu r). 
\ee
The Bessel function $K_0(\mu r)$ vanishes rapidly 
at infinity. Here, and hence also 
in the Einstein and Cotton tensors, the mass and spin contributions
become intertwined; nevertheless, inserting this 
metric into our expression (13) 
indeed gives the desired answer. [This is rather obvious, since 
we are using precisely the solution whose right-hand side is a 
sum of localized mass 
$T_{00}$ and spin $T_{0i}$ sources, and the corresponding charges
are defined to be their integrals, in this weak-field system.
The spin's expression has the same form as the energy's, 
but in terms of the rotational Killing vector.

In summary, we have constructed the conserved charges, particularly 
the energy and spin, in cosmological TMG for the relevant 
asymptotically constant curvature backgrounds.
Our formula is generic, applicable to both types of solutions as 
seen by explicit insertion of asymptotic metric expressions. In 
particular, we have also demonstrated that the energy of SdS  
does not receive
any contribution from the Cotton tensor, extending the 
$\Lambda =0$ result of \cite{deser2}

This research was supported by
NSF Grant 99-73935.


\end{document}